\documentclass[aps,pre,reprint,amsmath,superscriptaddress]{revtex4-1}
\usepackage{graphicx}
\usepackage{hyperref}
\usepackage{color}
\begin{document}
\title{Non-Close-Packed Three-Dimensional Quasicrystals}

\author{Pablo F.\ Damasceno}
\email{pablo.damasceno@ucsf.edu}
\affiliation{Applied Physics Program, University of Michigan, Ann Arbor, Michigan 48109, USA.}
\affiliation{Department of Cellular and Molecular Pharmacology, University of California San Francisco, San Francisco, CA 94158, USA}

\author{Sharon C.\ Glotzer}
\email{sglotzer@umich.edu}
\affiliation{Applied Physics Program, University of Michigan, Ann Arbor, Michigan 48109, USA.}
\affiliation{Department of Materials Science and Engineering, University of Michigan, Ann Arbor, Michigan 48109, USA}
\affiliation{Department of Chemical Engineering, University of Michigan, Ann Arbor, Michigan 48109, USA}

\author{Michael Engel}
\email{michael.engel@fau.de}
\affiliation{Department of Chemical Engineering, University of Michigan, Ann Arbor, Michigan 48109, USA}
\affiliation{Institute for Multiscale Simulation, Friedrich-Alexander-University Erlangen-Nuremberg, 91052 Erlangen, Germany}

\date{\today}

\begin{abstract}
Quasicrystals are frequently encountered in condensed matter.
They are important candidates for equilibrium phases from the atomic scale to the nanoscale.
Here, we investigate the computational self-assembly of four quasicrystals in a single model system of identical particles interacting with a tunable isotropic pair potential.
We reproduce a known icosahedral quasicrystal and report a decagonal quasicrystal, a dodecagonal quasicrystal, and an octagonal quasicrystal.
The quasicrystals have low coordination number or occur in systems with mesoscale density variations.
We also report a network gel phase.
\end{abstract}
\pacs{asa}
\maketitle


\section{Introduction}

The history of quasicrystals is characterized by three waves of discoveries: (i)~intermetallic quasicrystals, (ii)~molecular and nanoscale quasicrystals, and (iii)~quasicrystals in computer simulations.
The first wave started with the report of an icosahedral phase in Al-(Mn,Fe,Cr) alloys in 1984~\cite{Shechtman1984}.
Within four years, a dodecagonal phase in Ni-Cr alloys~\cite{Ishimasa1985}, a decagonal phase in Al-Mn alloys~\cite{Bendersky1985}, and an octagonal phase in (V,Cr)-Ni-Si alloys~\cite{Wang1987} followed.
These early intermetallic quasicrystals have in common that they are metastable grains in rapidly cooled alloys.
Thermodynamically stable quasicrystals were found later~\cite{Steurer2009,Tsai2013}.
All of them are densely packed crystal structures.

With the advent of molecular and nanoscale self-assembly, dodecagonal quasicrystals were reported in various systems, such as dendrimeric macromolecules in 2004~\cite{Zeng2004}, star polymers~\cite{Takano2005}, nanoparticles~\cite{Talapin2009,Ye2017}, mesoporous silica~\cite{Xiao2012}, perovskite thin films~\cite{Forster2013}, and coordination networks~\cite{Urgel2016}, as well as decagonal quasicrystals of pentameric molecules~\cite{Wasio2014}, and octakaidecagonal quasicrystals of block co-polymer micelles~\cite{Fischer2011}.
While the decoration of quasiperiodic tilings varies depending on the choice of building block (atoms, molecules, nanoparticles), the tilings themselves and thus their symmetries are universal across length scales.
These and other similarities suggest there are mechanisms for quasicrystal formation and stabilization that are independent of the identity of the building blocks.
Computer simulations can then be a valuable tool to supplement experimental observations.
Of particular importance are simulations that capture essential features common to all condensed matter systems forming quasicrystals.
Such simulations of model systems have the advantage that they are fast and require the tuning of only a few parameters.

Most early computer simulations reporting quasicrystals were conducted in two dimensions~\cite{Engel2008,Dotera2012}, either with lattice models~\cite{Tang1990,Oxborrow1993} or with freely moving particles~\cite{Lancon1986,Widom1987}.
Two-dimensional quasicrystals are now known in many models~\cite{Jagla1998,Quandt1999,Skibinsky1999,Engel2007,Schmiedeberg2008,Barkan2014,Dotera2014,Zu2017}.
As sources for their stabilization either competing length scales~\cite{Engel2007,Lifshitz2007,Barkan2011,VanderLinden2012,Barkan2014,Dotera2014} or appropriate bond angles~\cite{VanderLinden2012,Reinhardt2013,Reinhardt2017} have been identified, though neither is strictly necessary~\cite{Zu2017}.
Both approaches introduce local order that is incompatible with periodic lattices, like for example pentagonal rings of five particles.
Phase diagrams of soft-core systems~\cite{Dotera2014,Pattabhiraman2015,Schoberth2016,Pattabhiraman2017} and the effect of quasiperiodicity on particle dynamics~\cite{Engel2010,Hielscher2017} and stabilization~\cite{Kiselev2012} have been studied in detail.
Another direction are continuum simulations, such as phase field models~\cite{Archer2013,Achim2014,Archer2015,Jiang2015}.
In all cases, quasicrystals formed spontaneously from disordered starting configurations.

Computational discovery of quasicrystals in three dimensions is much less developed.
Because there is no general strategy to search for quasicrystals, observations are often accidental or require trial-and-error.
That is the case for the work of Dzugutov in 1993, who stumbled upon a Frank-Kasper-type dodecagonal quasicrystal~\cite{Dzugutov1993} while investigating a simple monatomic liquid with icosahedral inherent local order~\cite{Dzugutov1992}.
Computational studies of dodecagonal quasicrystals comprise a polymeric alloy~\cite{Dotera2006}, hard polyhedra~\cite{Haji-Akbari2009,Haji-Akbari2011,Damasceno2012}, confined bilayer water~\cite{Johnston2010} and silicon~\cite{Johnston2011}, deformable micelles and micelles self-assembled from both mono- and di-tethered nanoparticles~\cite{Iacovella2011}, smectic layers~\cite{Metere2016}, and polymer-tethered POSS cubes~\cite{Yue2016}.
Due to the complexity of self-assembly simulations in three dimensions and the apparent abundance of dodecagonal quasicrystals~\cite{Barkan2011,Dotera2011}, other quasicrystal symmetries received much less attention.

The list of quasicrystal simulations in three dimensions with symmetries other than dodecagonal is relatively short.
Phases resembling binary icosahedral~\cite{Roth1995} and decagonal~\cite{Roth1997} quasicrystals were investigated early on, but the system sizes were too small for unambiguous structure identification.
In other studies of icosahedral crystals particles were pinned on a lattice~\cite{Dmitrienko1995, Hann2016}, which confirmed that quasicrystals can grow efficiently.
Reliable simulation of non-dodecagonal quasicrystal growth was achieved only recently.
The exhaustive list includes an icosahedral quasicrystal with an oscillatory pair potential~\cite{Engel2015}, the observation of a similar icosahedral quasicrystal in a phase field model~\cite{Subramanian2016}, and the finding of a decagonal quasicrystal in a core/shell model~\cite{Ryltsev2015}.
We also mention the proposal of an octagonal quasicrystal related to $\beta$-Mn~\cite{Elenius2009}, which, however, has not yet been observed in simulation.

In this contribution, we report three axial quasicrystals together with the icosahedral quasicrystal from earlier work~\cite{Engel2015}.
The four quasicrystals are observed with a tunable pair potential that was originally intended to stabilize a diamond lattice.
The axial quasicrystals are rather unusual.
The decagonal quasicrystal contains pentagonal spirals with handedness that break ten-fold rotational symmetry.
The dodecagonal quasicrystal consists of narrow lamellae with hexagonal crystal structures in two layers that are rotated relative to one another by multiples of $30^\circ$.
Finally, the octagonal quasicrystal has a relationship to $\beta$-Mn but incorporates pairs of helices of equal handedness at the location of the tile vertices.
All quasicrystals form robustly in molecular dynamics simulations and can be grown as single crystals.


\section{Methods}

Our model system consists of identical particles interacting via an isotropic pair potential.
We varied the functional form of the pair potential, temperature, and pressure.
This is a simple simulation setup that can easily be investigated with any standard molecular dynamics code.

\subsection{Pair potential design}

Pair potentials were constructed by a series of educated guesses.
Our initial aim was to search for assemblies in parameter space in the vicinity of the diamond crystal. This is a good choice because tetrahedral coordination has proven to be promising for stabilizing quasicrystals~\cite{Haji-Akbari2009,Engel2015}.
The fact that diamond is an open structure with space for diffusion and reorganization helps to avoid kinetic traps.

We started by constructing a tabulated potential of mean force $V_\text{PMF}(r)=-k_BT_0\ln(g(r))$ from the radial distribution function $g(r)$ of a known diamond-forming system.
Configuration data for diamond was taken from earlier work on hard truncated tetrahedra~\cite{Damasceno2012} at temperatures close to but below the melting temperature.
By construction, $V_\text{PMF}$ has oscillations that resemble those in the oscillatory pair potential~\cite{Mihalkovic2012} and Friedel oscillations in metals~\cite{Friedel1952,Dzugutov1993}.
It is therefore a pair potential that is similar to those typically used to study alloys.
In the following we measure energy in units of $k_BT_0$ and temperature in units of $T_0$ by setting $k_BT_0=T_0=1$.

\begin{figure}
\centering
\includegraphics[width=\columnwidth]{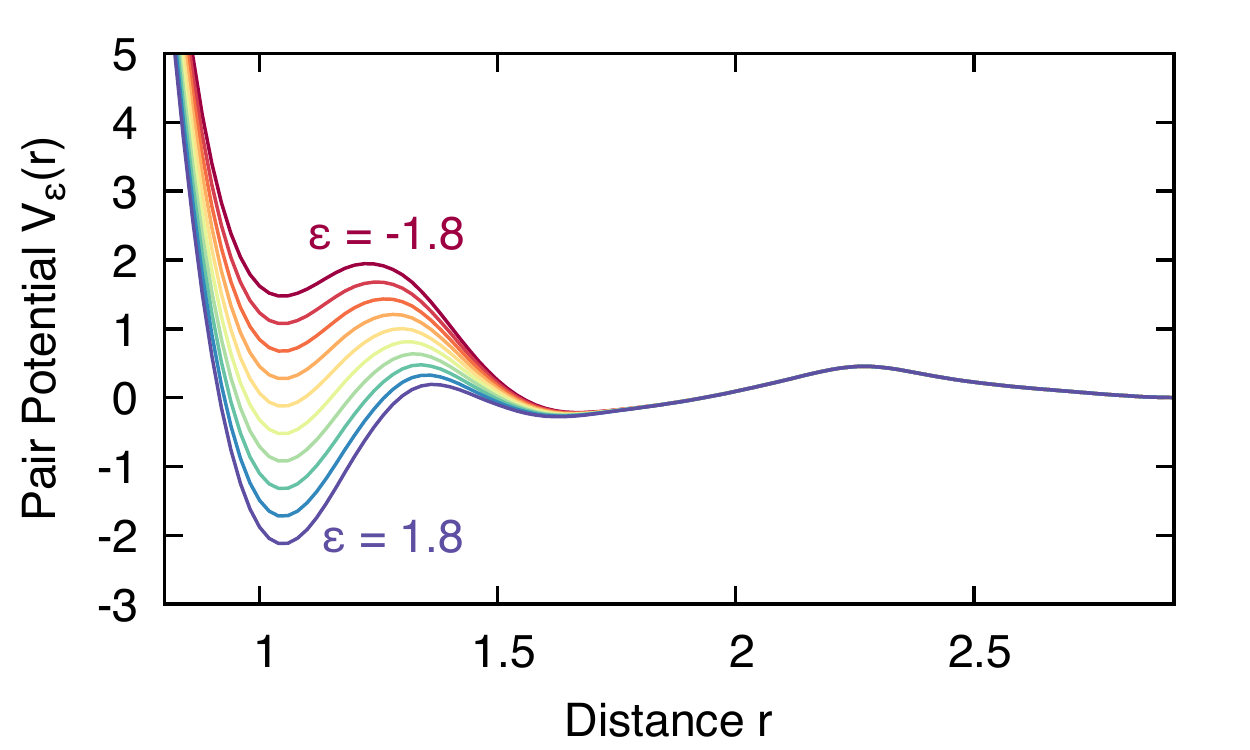}
\caption{
The family of isotropic pair potentials $V_\epsilon(r)$ used in this work.
The potentials are derived from the potential of mean force of a diamond crystal.
The ratio of the positions of the first and second well of the potential is close to the edge length to circumsphere radius ratio of a tetrahedron, $\sqrt{8/3}\approx1.63$, as required for the stabilization of local tetrahedral order.
The second well is wide, which introduces flexibility in the preferred bond angle and allows the potential to stabilize a variety of different crystal structures.
Depending on the choice of $\epsilon$, the first minimum can be higher or lower than the second minimum.
}
\label{fig1}
\end{figure}

Next, we created a family of pair potentials by adding to $V_\text{PMF}$ a Gaussian located at the first minimum $r_1$ with width $\sigma$ chosen equal to the width of the first well,
\begin{equation}
V_\epsilon(r)=V_\text{PMF}(r)-\epsilon\exp\left(-\frac{(r-r_1)^2}{2\sigma^2}\right)-V(r_3).
\end{equation}
We cut off the potential in the third minimum $r_3=2.94$ where the potential is shifted to be zero.
The motivation to add a Gaussian was to control the coordination number by varying the depth of the first well while, at the same time, keeping bond angles unchanged by not varying the ratio between well positions~\cite{Engel2015}.
Overall, ten smooth pair potentials with $\epsilon=-2.2 + 0.4 n$, $n=1,\ldots,10$ were constructed (Fig.~\ref{fig1}).
Data tables of the ten potentials are included as Supplementary Information.

\subsection{Self-assembly simulations}

Molecular dynamics simulations were performed with HOOMD-blue~\cite{Anderson2008,Glaser2015} in the $NpT$ ensemble using a cubic box with periodic boundary conditions.
In total $30\times10$ self-assembly simulations were initialized for the ten pair potentials and pressures in the range $1.1\leq p \leq4.0$ by placing $N=50000$ particles randomly in the box and lowering the temperature linearly from $T_\text{init}$ to $T_\text{end}$ over 120 million molecular dynamics time steps.
The initial temperature $T_\text{init}$ was chosen in the fluid regime above the melting temperature and the final temperature $T_\text{end}$ well below the melting temperature in the solid.
Those values were found prior to production runs by performing smaller simulations with $N=4096$ particles and lowering the temperature linearly from $5.0$ to $0.1$ over 100 million molecular dynamics time steps.

The final frames of the self-assembly simulations were analyzed without applying a quench to zero temperature or thermal averaging.
Crystal structures were identified manually with the help of diffraction patterns and bond-orientational order diagrams following literature procedures~\cite{Engel2015}.
In situations with several (quasi-)crystalline grains, we focused on the dominant phase.
Particles are drawn in figures with diameter equal to the nearest-neighbor distance if the whole system is shown, or with diameter equal to 40\% of their nearest-neighbor distance if projections or small parts of the system are shown.
Bond-orientational order diagrams are shown in stereographic projection.


\section{Results and discussion}

\begin{figure}
\centering
\includegraphics[width=0.85\columnwidth]{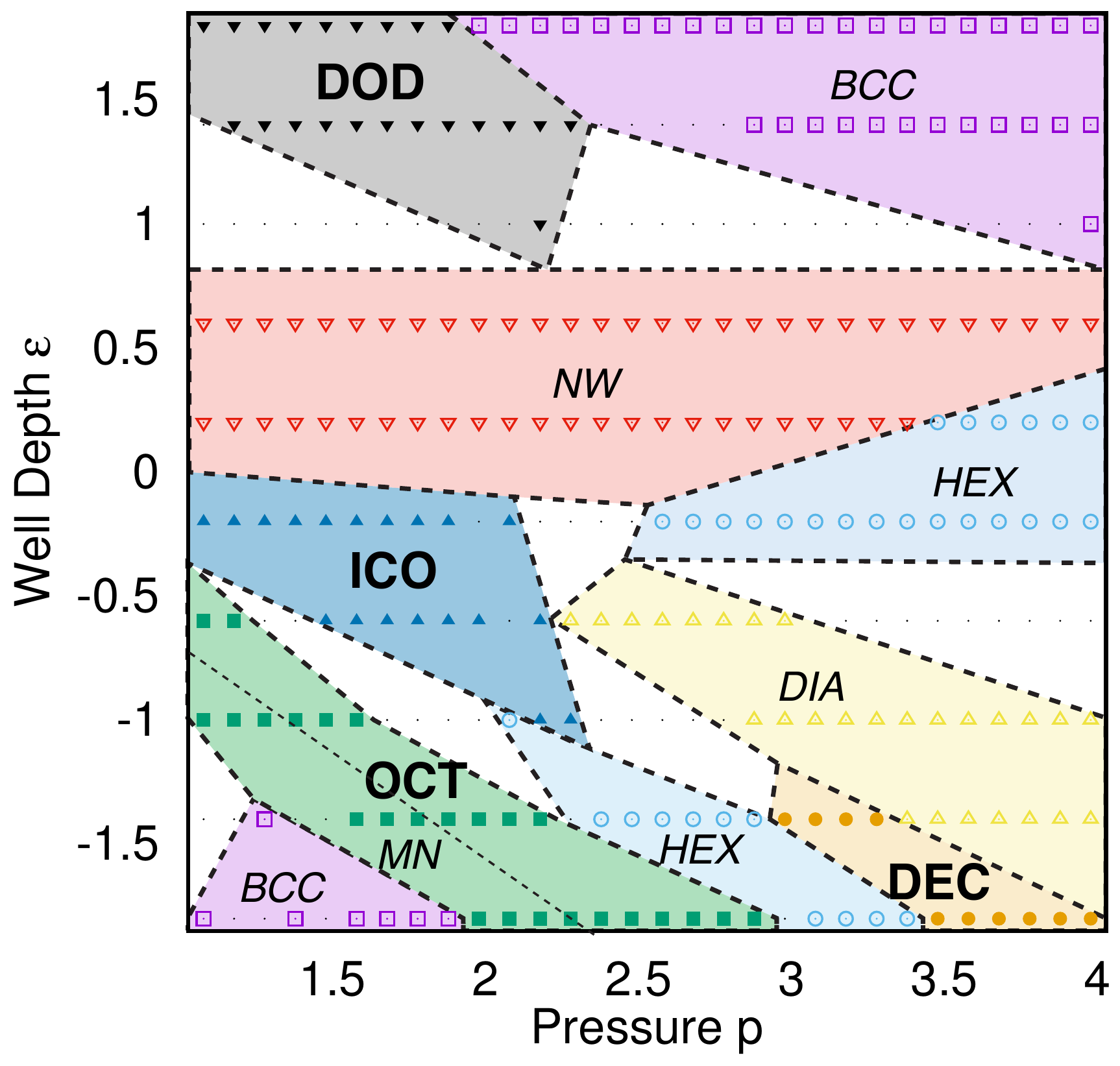}
\caption{
Phase observation diagram in the pressure--well depth plane for the isotropic pair potentials in Fig.~\ref{fig1}.
Simulations were initialized in a disordered fluid phase at high temperature and slowly cooled at fixed pressure until self-assembly was observed.
Each colored region highlights an area where a certain phase is predominantly observed in simulation.
Labels are defined in the text.
Bold labels specify phases with quasiperiodic order.
White areas correspond to simulations that remain disordered, homogeneous glasses on the timescale of our runs.
}
\label{fig2}
\end{figure}

\begin{figure}
\centering
\includegraphics[width=0.9\columnwidth]{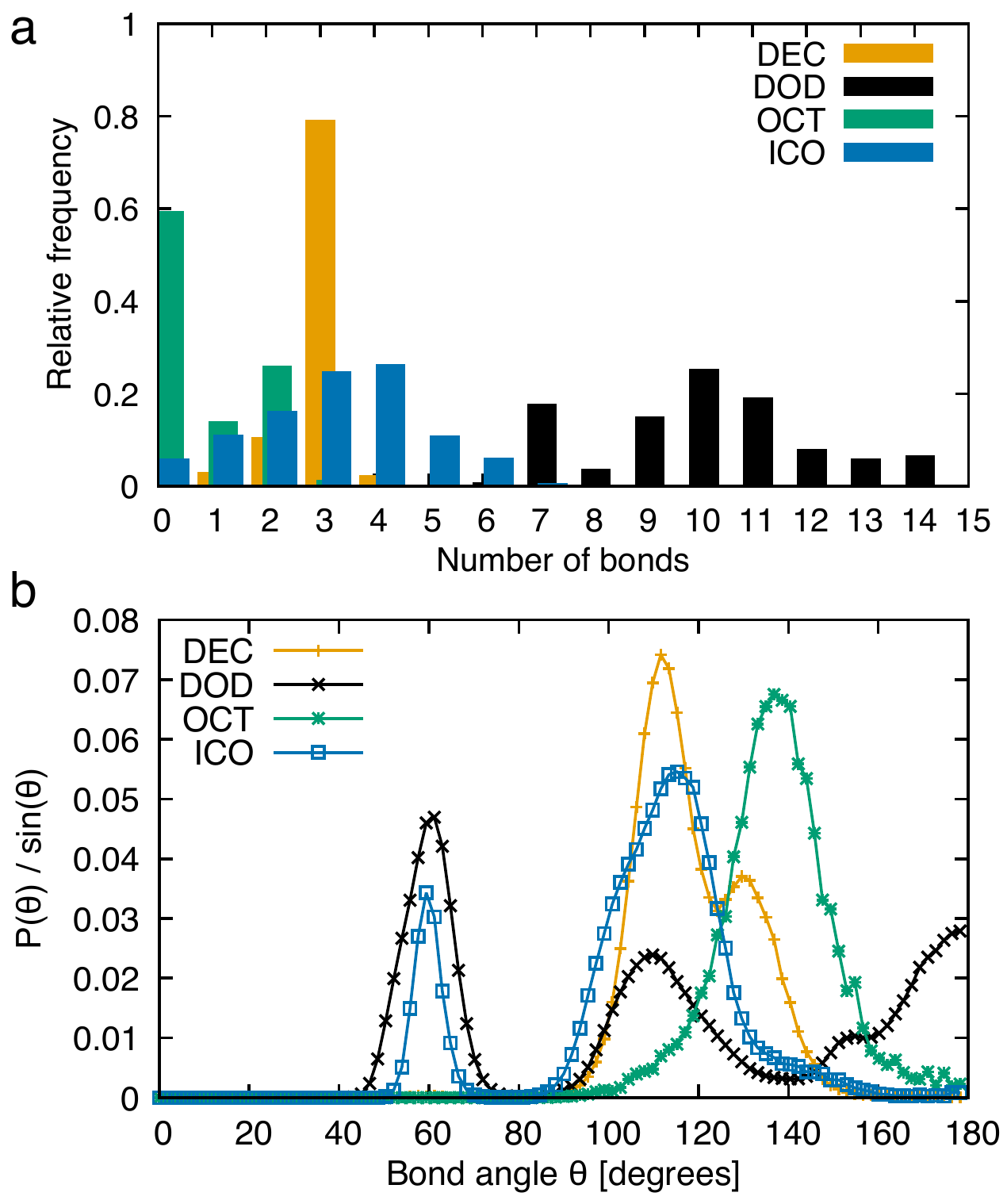}
\caption{
(a)~Probability histogram for a particle to have a certain number of nearest neighbors.
Nearest neighbors are defined as those corresponding to the first peak in the radial distribution function.
(b)~Distribution of the angles between nearest-neighbor bonds.
The probability of a given angle to occur $P(\theta)$ is divided by the Jacobian $\sin(\theta)$.
Data is shown for the four quasicrystals at parameters given in the captions of Figs.~\ref{fig4}-\ref{fig7}.
}
\label{fig3}
\end{figure}

The crystal structures found in the self-assembly simulations are summarized in a phase observation diagram (Fig.~\ref{fig2}).
Overall, 70\% of the simulations developed some form of (quasi-)crystalline order, a number comparable to simulations with similar potentials~\cite{Engel2015}.
We observe eight ordered phases, among them four quasicrystals and four periodic crystals.

\begin{figure*}
\centering
\includegraphics[width=0.8\textwidth]{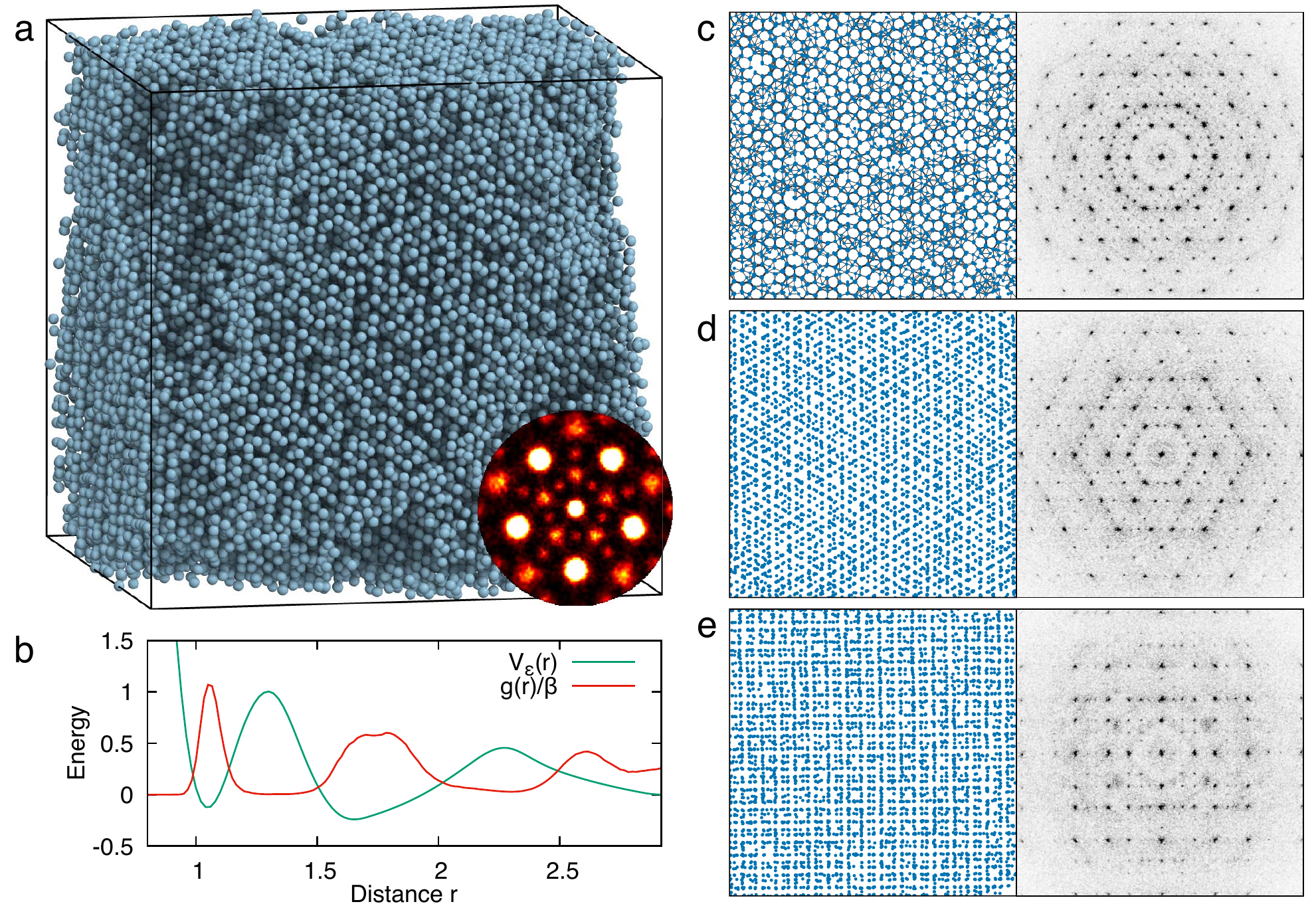}
\caption{
Icosahedral quasicrystal for the parameters $\epsilon=-0.2$, $p=1.5$. 
(a)~Final frame of the self-assembly simulation at $T=0.2$.
A single-crystalline grain is shown.
The inset shows the bond-orientational order diagram.
(b)~Pair potential $V_\epsilon(r)$ and scaled radial distribution function $\beta^{-1}g(r)$ with $\beta^{-1}=k_\text{B}T$.
(c-e)~Snapshots (left) and corresponding diffraction images (right) along a five-fold~(c), three-fold~(d), and two-fold~(e) axis.
}
\label{fig4}
\end{figure*}

The periodic crystals correspond to the well-known body-centered cubic (BCC), diamond (DIA) and $\beta$-Mn (MN) structures, which are frequently observed in simulations of identical particles~\cite{Rechtsman2007,Elenius2009,Damasceno2012,Jain2013}.
We also observe a hexagonal phase (HEX) with Pearson symbol hP2~\cite{Engel2015}.
The quasicrystals include a decagonal quasicrystal (DEC), a dodecagonal quasicrystal (DOD), an octagonal quasicrystal (OCT), and an icosahedral quasicrystal (ICO).
All of them are non-close-packed with average coordination number well below twelve (Fig.~\ref{fig3}).
We also observe a gel-like disordered network phase (NW) that is distinct from glassy configurations usually occurring in simulations of spherical particles with short-range interactions.
The structures of the four quasicrystals and the network gel are now discussed in more detail.

\subsection{Icosahedral quasicrystal}

Icosahedral quasicrystals are quasiperiodic in all directions.
We reproduce a recently reported~\cite{Engel2015} icosahedral quasicrystal, ICO, in our system (Fig.~\ref{fig4}(a)).
The agreement of the present ICO with prior work is confirmed by comparing radial distribution functions, bond-orientational order diagrams, and diffraction patterns (Fig.~\ref{fig4}(b-e)).

ICO is robust under mild variations of the formation conditions and nucleates and grows easily in simulation, comparable to much simpler crystal phases.
Its coordination spans a broad range from zero to six neighbors (Fig.~\ref{fig3}(a)) with average coordination number 3.1, which is even lower coordinated than the low-density icosahedral quasicrystal reported previously~\cite{Engel2015}.
A comparison of ICO at different pressures shows that the structure is surprisingly flexible and responds to density fluctuations by varying the average coordination number over a broad range from 2.5 and 4.0.
Together, these two observations indicate that the stabilization of ICO is not controlled by nearest-neighbor bonds alone.
Energetic contributions stemming from second-neighbors are at least equally important because $V_\epsilon(r)$ is deeper at the second-neighbor shell, $1.5<r<2.0$, than at the nearest-neighbor shell, $0.9<r<1.2$ (Fig.~\ref{fig4}(b)).
The width of the second well $V_\epsilon(r)$ also helps to stabilize ICO because the second-neighbor shell is broad.
The distribution of nearest-neighbor angles shows peaks at the three-fold angles $60^\circ$ and $120^\circ$ and at the tetrahedral angle $109^\circ$ (Fig.~\ref{fig3}(b)).

\subsection{Decagonal quasicrystal}

\begin{figure}
\centering
\includegraphics[width=\columnwidth]{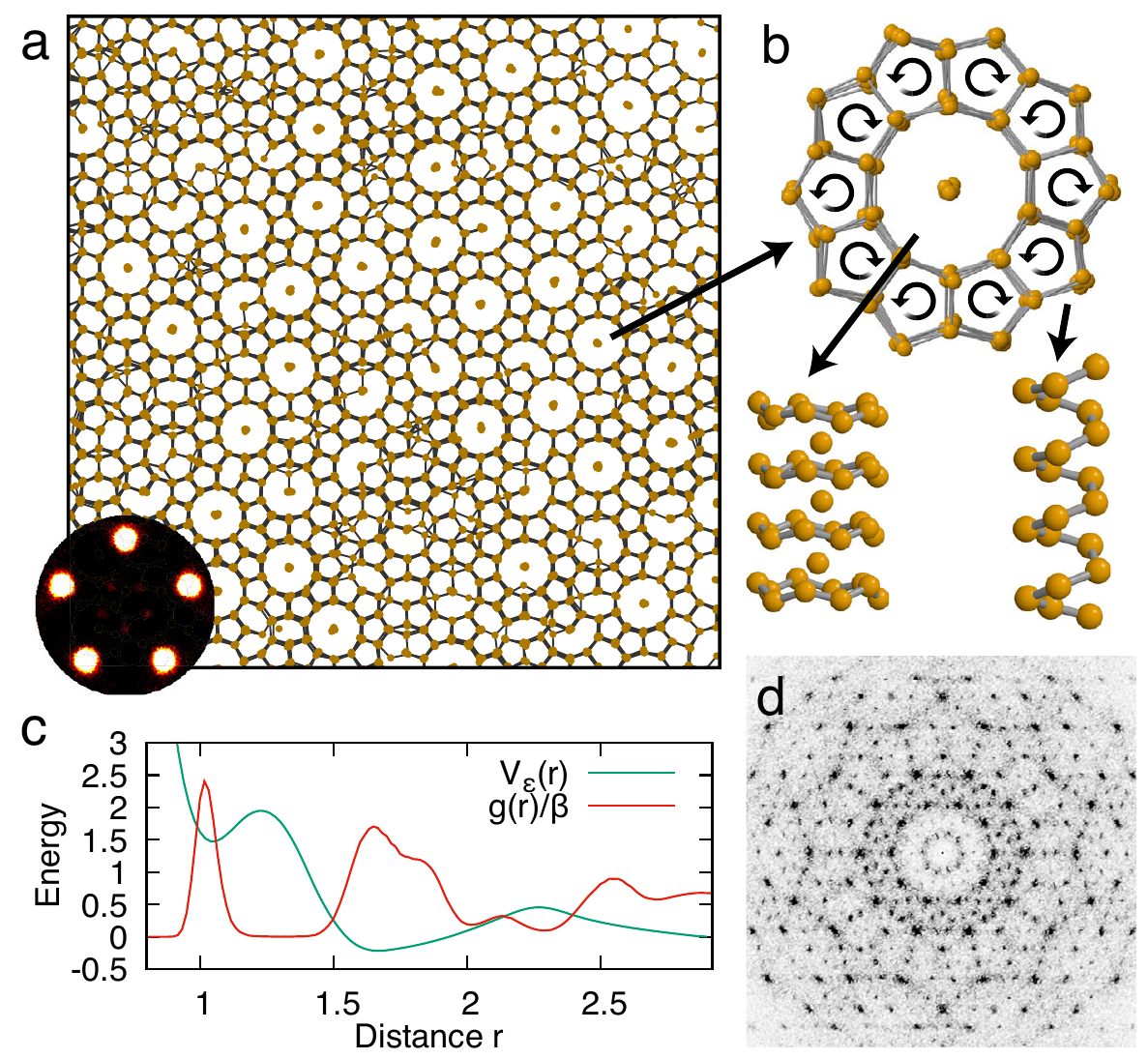}
\caption{
Decagonal quasicrystal for the parameters $\epsilon=-1.8$, $p=3.9$. 
(a)~Final frame of the self-assembly simulation at $T=0.2$ viewed along the ten-fold axis.
The inset shows the bond-orientational order diagram.
(b)~A decagon surrounded by ten pentagons is cut out.
Side views show that decagons are staggered and pentagons correspond to a helix.
(c)~Pair potential $V_\epsilon(r)$ and scaled radial distribution function $\beta^{-1}g(r)$.
(d)~Diffraction image along the ten-fold axis.
}
\label{fig5}
\end{figure}

Decagonal, dodecagonal, and octagonal quasicrystals are axial quasicrystals.
This means they are quasiperiodic in a plane and periodic perpendicular to this plane.
The decagonal quasicrystal we observe, DEC, is of different structure and of higher quality than the only other decagonal quasicrystal reported in simulation~\cite{Ryltsev2015}.
The most noticeable features of DEC are decagonal rings when viewed along the ten-fold axis (Fig.~\ref{fig5}(a)).
Whereas decagonal rings in~\cite{Ryltsev2015} are directly in contact, decagonal rings in DEC are separated by other tiles.
In projection, DEC is identical to a random T\"ubingen triangle tiling~\cite{Baake1990}, which is frequently observed in two-dimensional simulations~\cite{Jagla1998,Skibinsky1999,Engel2007,Dotera2012}.
Its three-dimensional extension is reported here for the first time.

A view from the side reveals that decagonal rings are not planar but staggered, just like the C$_6$-ring in cyclohexane (Fig.~\ref{fig5}(a)).
Pentagons, typically found next to decagonal rings in the tiling, correspond to helices with a pitch of five particles.
More generally, any even-membered ring (hexagon, decagon) is planar while any odd-membered ring (pentagon, nonagon) is a helix.
The handedness of a helix is determined by the orientation of its ring in the tiling projection.
For example, the handedness of pentagons alternates around the decagon.
As a result, DEC is not invariant under inversion and has a chiral five-dimensional space group~\cite{Steurer2009}.
The chirality of the space group is apparent also in the bond orientational order diagram, which exhibits five-fold rotational symmetry only.

The pair potential for DEC (Fig.~\ref{fig5}(c)) has a first well that is significantly higher than the second well.
This means DEC is stabilized mostly by second-neighbor interactions and pressure is required to push particles into the nearest-neighbor shell.
A majority of the particles in DEC have three nearest neighbors (Fig.~\ref{fig3}(a)).
Only a few particles (e.g.\ those in the centers of decagons) have lower coordination.
The most common bond angles (Fig.~\ref{fig3}(b)) are the tetrahedron angle $109^\circ$, found in pentagon helices, and $135^\circ$, found in decagonal rings.
As expected, the diffraction image of DEC (Fig.~\ref{fig5}(d)) is identical to the diffraction image of the T\"ubingen triangle tiling~\cite{Engel2007}.

\subsection{Dodecagonal quasicrystal}

\begin{figure}
\centering
\includegraphics[width=\columnwidth]{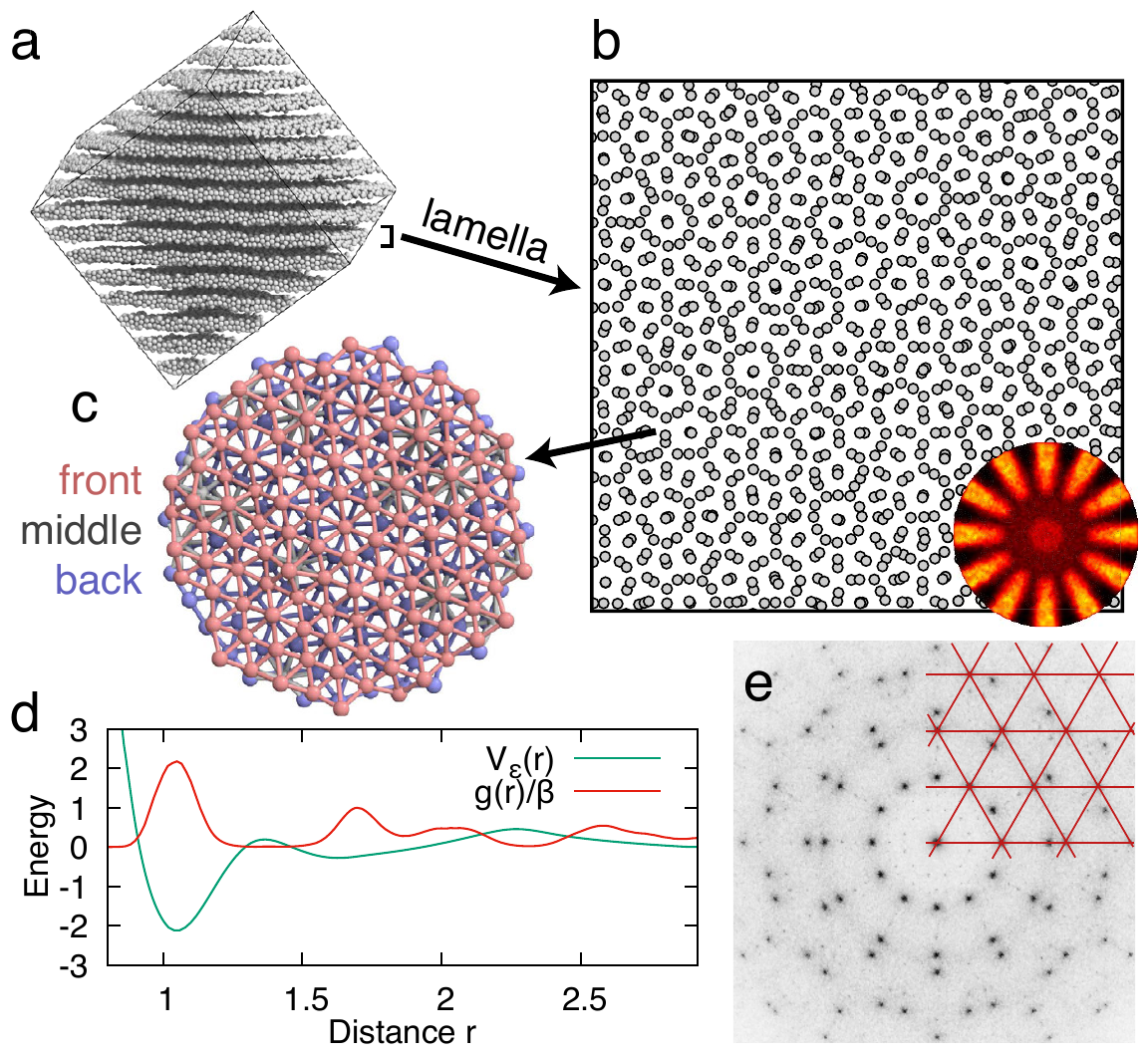}
\caption{
Dodecagonal quasicrystal for the parameters $\epsilon=1.8$, $p=1.2$. 
(a)~The final frame of the self-assembly simulation at $T=0.2$ viewed along the dodecagonal axis.
A lamellar phase is present.
(b)~A single lamella viewed along its normal axis has dodecagonal order.
The inset shows the bond-orientational order diagram.
(c)~The close-up view reveals that the lamella consists of two rotated hexagonal layers of particles (red and blue).
A few particles (gray) are in-between the two hexagonal layers.
(d)~Pair potential $V_\epsilon(r)$ and scaled radial distribution function $\beta^{-1}g(r)$.
(e)~The diffraction image along the twelve-fold axis is a sum of two hexagonal reciprocal lattices.
A part of one hexagonal lattice is shown as a red overlay.
}
\label{fig6}
\end{figure}

As discussed in the introduction, dodecagonal quasicrystals are frequently found in soft matter experiments~\cite{Zeng2004,Takano2005,Talapin2009,Ye2017,Xiao2012,Forster2013,Urgel2016,Fischer2011} and simulations~\cite{Dzugutov1993,Metere2016}.
The dodecagonal quasicrystal observed here, DOD, shows similarity with a recently reported smectic dodecagonal quasicrystal~\cite{Metere2016}.
Both consist of hexagonal layers rotated relative to one another by $30^\circ$ and not all in direct contact.
However, whereas the smectic dodecagonal quasicrystal has all layers spaced equally along the twelve-fold axis, our system has thicker lamellae (Fig.~\ref{fig6}(a)).
Similar rotations of hexagonal layers have been reported in a micellar colloidal dodecagonal quasicrystal~\cite{Fischer2011}.

Each individual lamella in DOD has dodecagonal order (Fig.~\ref{fig6}(b)).
Dodecagonal order is synchronized between lamellae, which means it extends throughout the whole system.
However, synchronization is weak.
We cannot exclude that periodic boundaries assisted in the synchronization by introducing correlations.
Boundary effects could play a role because, in contrast to the other crystals and quasicrystals we observe, lamellae are relatively far apart and therefore only weekly coupled.
Each lamella consists of two layers of particles that project on hexagonal lattices along the twelve-fold axis (Fig.~\ref{fig6}(c)).
This is confirmed by the diffraction image (Fig.~\ref{fig6}(e)), which is the superposition of diffraction images of hexagonal lattices rotated by $30^\circ$.
The hexagonal layers are not flat but exhibit some puckering.
Additional particles are sometimes introduced between the layers but rarely do we find more than a few of them.

The pair potential for DOD (Fig.~\ref{fig6}(d)) has a deep first-neighbor well that dominates the interaction.
Additional peaks destabilize close-packing (e.g.\ face-centered cubic) and cause the appearance of lamellae.
Particles have a range of coordination numbers (Fig.~\ref{fig3}(a)) depending on whether a particle sits deep within the layer or is more exposed at the surface.
Coordination is overall much higher than for the other quasicrystals due to the deeper first well.
Common bond angles are $60^\circ$, $109^\circ$, $155^\circ$, and $180^\circ$ (Fig.~\ref{fig3}(b)).

We did not search for the existence of a lamellar liquid crystalline phase with freely mobile particles in each lamella.
But we expect it to appear between DOD and the isotropic liquid for appropriately chosen parameters.

\subsection{Octagonal quasicrystal}

\begin{figure}
\centering
\includegraphics[width=\columnwidth]{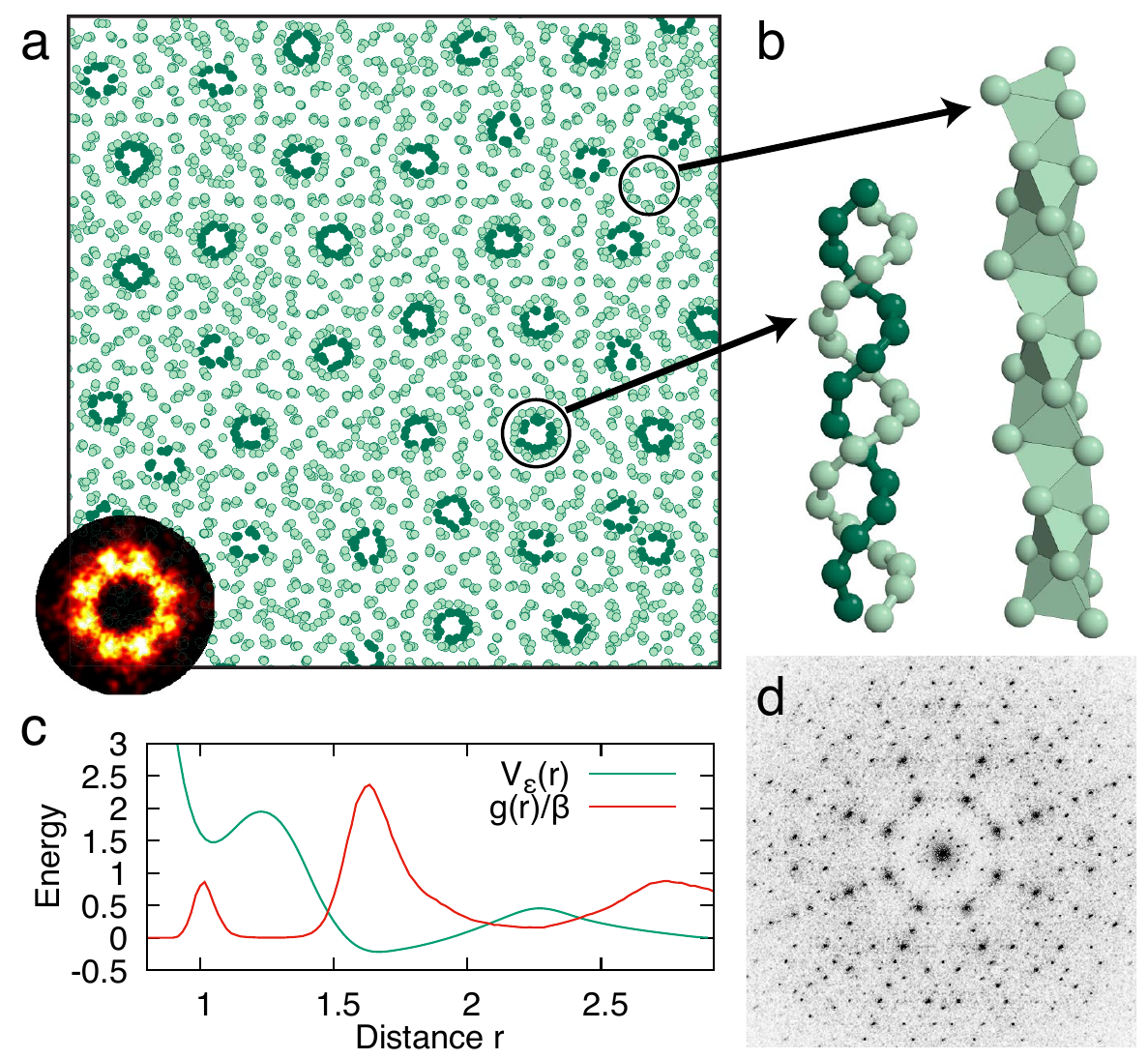}
\caption{
Octagonal quasicrystal for the parameters $\epsilon=-1.8$, $p=2.5$. 
(a)~Final frame of the self-assembly simulation at $T=0.2$ viewed along the eight-fold axis.
Some particles form double rings, shown in dark green for the inner ring and light green for the outer ring.
Color is for highlighting purposes only.
(b)~A side view of a double ring reveals that it is a pair of helices.
In the absence of defects all pairs of helices have the same handedness.
Rings of light green particles without inner rings are slightly deformed tetrahelices.
(c)~Pair potential $V_\epsilon(r)$ and scaled radial distribution function $\beta^{-1}g(r)$.
(d)~Diffraction image along the eight-fold axis.
}
\label{fig7}
\end{figure}

Besides this work, octagonal quasicrystals have not been observed in simulation.
The octagonal quasicrystal in our data, OCT, is obtained with the same pair potential as DEC but at lower pressure.
Particles align in planes with normals pointing in four directions perpendicular to an eight-fold axis (Fig.~\ref{fig7}(a)).
A characteristic feature is the presence of double rings along the eight-fold axis.
The double rings are spaced at discrete distances and with angles multiples of $45^\circ$ from one another, forming the vertices of a quasiperiodic tiling with octahedral symmetry.

When viewed from the side (Fig.~\ref{fig7}(b)), the double rings are revealed to be a pair of helices with equal handedness.
The inner ring has a pitch length of six particles and the outer ring has a pitch length of eight particles.
The combination of the difference in pitch lengths and the difference in radii allows equal bond length in both helices.
All pairs of helices in double rings have equal handedness throughout the quasicrystal, although we sometimes find defects that flip the handedness.
This means OCT is not invariant under inversion and has a chiral five-dimensional space group~\cite{Steurer2009}.
We also find wider rings in the structure.
They correspond to slightly deformed tetrahelices as apparent when viewed from the side.
The diffraction image (Fig.~\ref{fig7}(d)) and radial distribution function of OCT show a similarity with a proposed octagonal quasicrystal based on the $\beta$-Mn crystal structure~\cite{Elenius2009}.
However, the elementary square and rhomb tiles that can be extracted from our simulation data are larger by a factor of $\sqrt{2}$.
In our simulations, OCT competes with $\beta$-Mn and it is not possible to separate them clearly in the phase observation diagram (Fig.~\ref{fig1}).
However, we observe a preference for OCT towards higher pressure.

The pair potential for OCT has a high first well (Fig.~\ref{fig7}(c)).
Because the applied pressure is low, there are only a few nearest-neighbor bonds.
These bonds occur exclusively within double rings, which means the majority of the particles do not have any nearest-neighbor bonds (Fig.~\ref{fig3}(a)).
We are not aware of an experimental system with such a low coordination number.
The bond angle distribution shows a single peak around $140^\circ$ (Fig.~\ref{fig3}(b)), which is the angle formed along the pair of helices.

\subsection{Network gels}

\begin{figure*}
\centering
\includegraphics[width=0.8\textwidth]{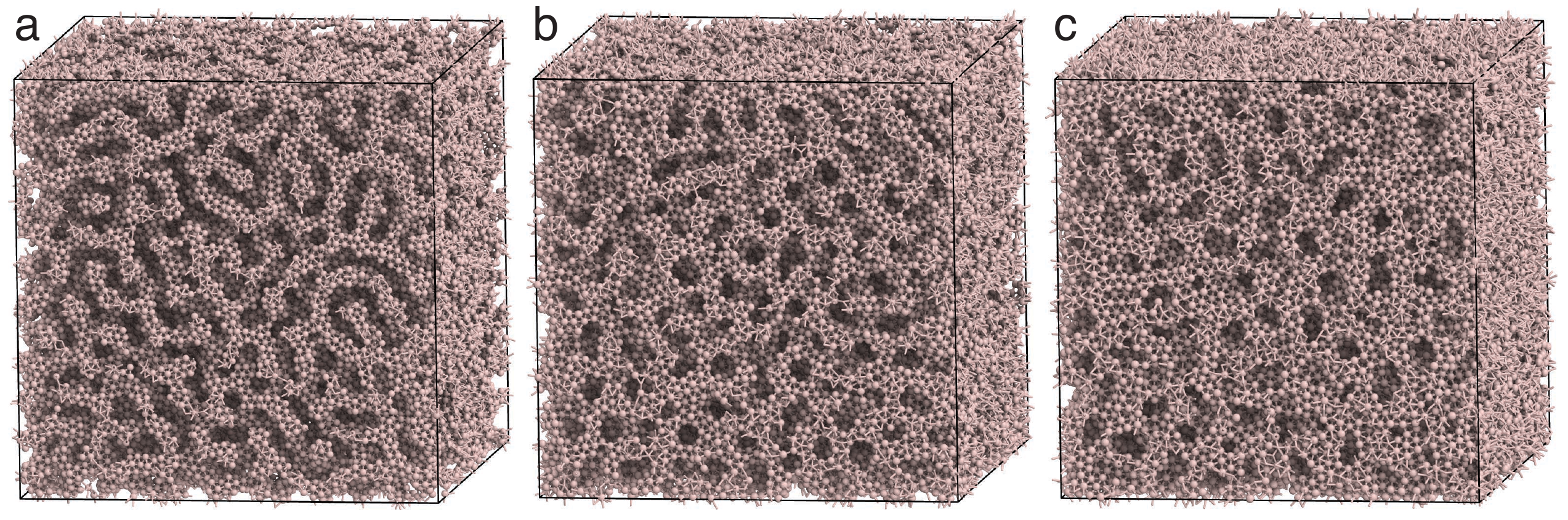}
\caption{
Network-like gels for the parameters $\epsilon=0.6$, (a)~$p=1.1$, (b)~$p=2.5$, and (c)~$p=4.0$. 
The gels are disordered and porous. The number of pores, but not their average lateral diameter, decreases with pressure.
}
\label{fig8}
\end{figure*}

Inhomogeneous clustering is observed for pair potentials with an intermediate depth of the first well.
The system develops mesoscale order under such a condition without the need for short-range order or long-range order.
We observe in our simulations gel-like networks of densely packed particles separated by empty pores (Fig.~\ref{fig8}).
Similar inhomogeneous networks frequently appear with isotropic pair potentials if the potential is soft or has competing length scales~\cite{Malescio2003,Camp2003,Sciortino2004,Glaser2007,Elenius2009a,Edlund2010}.

It is unclear if the network gels will order at lower temperature.
However, DOD appears to be one limit that occurs when the pores are large enough to aggregate and form lamellae.
Other mesophases including columnar phases or cluster phases~\cite{Mladek2006} might appear nearby in parameter space.


\section{Conclusion}

This work reported three axial quasicrystals in three dimensions and investigated an icosahedral quasicrystal using molecular dynamics simulations.
The quasicrystals are distinct from quasicrystals found in alloys and distinct from quasicrystals known in molecular and nanoscale matter as particles have unusual coordination that is not close-packed.
Stabilization of the quasicrystals observed in our simulations requires the interplay of two length scales, which might be non-trivial in experiments.
Nevertheless, our model system is well suited to study structure formation and other aspects of quasicrystals because particle dynamics and quasicrystal growth are fast.
Future work could extend our computational approach to binary systems that are more similar to alloys.
But even without such an extension, it is possible to consider our simulations as a coarse-grained description of experiments by identifying particles in simulation not with individual atoms or molecules but with groups of atoms or molecules.
The observation of network gels with inhomogeneous density variations also provides opportunities for follow-up work.

Our search for quasicrystals in the vicinity, in parameter space, of the diamond crystal was more successful than expected.
The fact that we self-assembled the diamond crystal structure with an only slightly modified potential of mean force (a slightly deeper first well), an achievement that is not trivial or expected, suggests that the potential of mean force approach employed here is also a promising starting point for the stabilization of other crystal structures with pair potentials.
At the same time the connection between local tetrahedral order and the appearance of quasiperiodicity, in the four variants we observe here, is not well understood.
In this first report we did not attempt to prove thermodynamic stability or obtain phase diagrams.
Instead, we focused on the geometric description of the quasicrystal phases.
Our findings show new ways to achieve quasiperiodic long-range order and demonstrate that different types of quasicrystals appear naturally as condensed matter phases in three dimensions.


\section*{Acknowledgements}

This material is based upon work supported in part by the U.S.\ Army Research Office under Grant Award No.\ W911NF-10-1-0518 and by a Simons Investigator award from the Simons Foundation to S.C.G.
P.F.D.\ acknowledges support from the University of Michigan Rackham Predoctoral Fellowship Program.
This research used the computational resources of the Oak Ridge Leadership Computing Facility, which is a DOE Office of Science User Facility supported under contract DE-AC05-00OR22725, and was supported in part through computational resources and services by Advanced Research Computing at the University of Michigan, Ann Arbor.
M.E.\ acknowledges funding by Deutsche Forschungsgemeinschaft through the Cluster of Excellence Engineering of Advanced Materials (EXC 315/2) and support from the Central Institute for Scientific Computing (ZISC) and the Interdisciplinary Center for Functional Particle Systems (IZ-FPS) at FAU Erlangen-Nuremberg.

\section*{References}
\bibliography{quasicrystals}

\end{document}